\documentstyle{ioplppt}
\input epsf
\begin{document}
\jl{1}
\title{Finite-size scaling corrections in two-dimensional
Ising and Potts ferromagnets}[FSS corrections in two-dimensional magnets]
\author{S L A de Queiroz\ftnote{1}{E-mail: sldq@if.ufrj.br}}

\address{Instituto de F\'\ii sica, UFRJ, Caixa Postal 68528, 
21945--970 Rio de Janeiro RJ, Brazil}

\begin{abstract}
Finite-size corrections to scaling of critical correlation lengths and
free energies of Ising and three-state Potts ferromagnets are analysed
by numerical methods,
on  strips of width $N$ sites of square, triangular and honeycomb
lattices.
Strong evidence is given that the amplitudes of the
``analytical'' correction terms, $N^{-2}$, are
identically zero for triangular-- and honeycomb Ising systems. For
Potts spins, our results are broadly consistent with this
lattice-dependent pattern of cancellations,
though for correlation lengths non-vanishing (albeit rather small)
amplitudes cannot be entirely ruled out.
\end{abstract}

\maketitle

\section{Introduction}
The systematic study of sub-dominant terms in scaling 
provides researchers with guidelines on how to extrapolate  e.g.
finite-size results
to the thermodynamic limit~\cite{dds,bn,barber,luck,ball}.
Also, it may in
itself bring out connections to underlying physical properties. These
latter may be universal, such as the
relationship between critical free-energy finite-width correction
and conformal anomaly of the corresponding universality class~\cite{bcn},
or otherwise relate to details of the system under
consideration~\cite{bdn88}. 
In the present work we investigate the possible existence of a link
between lattice structure and presence (or absence) of specific
subdominant terms in finite-size scaling. A numerical analysis is
made of correlation-length and free energy data at criticality,
as given by the largest eigenvalues of the transfer matrix
(TM)~\cite{nig},
for Ising and three-state Potts ferromagnets on strips of
square, triangular and honeycomb lattices with homogeneous, isotropic 
nearest-neighbour couplings, and periodic boundary
conditions (PBC) across. 

We recall the following results from conformal invariance~\cite{cardy}
for the critical spin-spin correlation length, $\xi_N$, and
(dimensionless) negative free energy per site, $f_N$, on strips of width
$N$ sites with PBC~\cite{bcn,bdn88,cardy2,pf84}:
\begin{equation}
N/\pi\xi_N = \eta + a_{\xi}N^{-\omega}+ b_{\xi}N^{-\omega_1} + \ldots
\label{eq:1}
\end{equation}
\begin{equation}
N^2(f_N- f_{\infty})= c\pi/6 +
a_{f}N^{-\omega} +b_{f}N^{-\omega_1} + \ldots
\label{eq:2}
\end{equation}
\noindent where $\eta$ is the
decay-of-correlations exponent and $c$ the conformal anomaly
number, with respective exact values $\eta =1/4$, $c=1/2$ (Ising);
$\eta = 4/15$, $c =4/5$
(three-state Potts)~\cite{bcn,cardy,cardy2}. 
In terms of the two largest eigenvalues $\Lambda_N^0$ and $\Lambda_N^1$
of the column-to-column TM, 
$\xi_N^{-1}=\zeta \ln(\Lambda_N^0/\Lambda_N^1)$; $N f_N = 
\zeta \ln \Lambda_N^0$; the factor
$\zeta$ is unity for the square lattice and,
in triangular or honeycomb geometries, corrects for the fact that the
physical length added upon each application of the TM differs from
one lattice spacing~\cite{pf84,bww90,bn93}. 

Studies of the operator content of conformally invariant
theories, and the  perturbation theory (for finite-size
systems, with $N^{-1}$ as the perturbation variable) for
the corresponding operator product expansion~\cite{cardynp}
have shown how, in two dimensions, the allowed values of the
exponents $\omega$, $\omega_1$ etc are related to the scaling dimensions
of the respective set of (model-specific) irrelevant operators. These
exponents are therefore universal, as the set of their allowed values is 
fixed for a given model. It must be noted that terms with $\omega=2$, the
so-called ``analytical'' corrections, are expected in any theory, as
they are related to the conformal block of the identity
operator~\cite{cardynp}. It has also been shown~\cite{car86,abb88}
that no first-order corrections are expected for ground-state energies,
the dominant terms being second-order; this will be particularly relevant
for the three-state Potts model, as seen below.

Many numerical studies of corrections to
scaling pertaining to two-dimensional classical (i.e. Ising, Potts
etc) systems have actually been carried out in their {\it
one}-dimensional {\it quantum}
counterparts~\cite{abb88,malte,vGRV,rein87}, by taking advantage of
well-known correspondences~\cite{fs78,kogut}. While this is expected
to have no effect on the determination of the universal
correction exponents,
the amplitudes $a_{\xi}$, $a_f$, $b_{\xi}$,
$b_f$ etc are generally believed to be non-universal, though e. g. in
certain quantum chain systems they display similar dependences 
(Privman-Fisher universality)~\cite{pf84} on the anisotropy
parameter~\cite{malte}. On the other hand, the relationship
between $(1+1)$-dimensional chains and two-dimensional spin
systems is such that the latter are necessarily located on a square
lattice. Thus, much less is known about corrections to scaling for
spins on, say, triangular or honeycomb symmetries than on their
square counterpart. So far, the only explicit
(published) reference to the connection between lattice symmetry and
corrections to scaling seems to be a remark on the fact that,
on a square lattice operators of spin $\pm 4$ will appear, 
giving rise to  $N^{-2}$ corrections~\cite{cardynp}.

Our main purpose here is to estimate, through numerical work, the first
(i.e. lowest-order) few amplitudes
and exponents as given in Equations~(\ref{eq:1}) and~(\ref{eq:2}), for
Ising and three-state Potts spins on square, triangular and
honeycomb geometries. We shall usually assume, for each model, 
definite values for the two or three exponents used in our fits,
taking our hints from conformal invariance 
theory~\cite{cardynp,car86,abb88}. 

In particular, we shall seek instances in which, for a given spin model 
and exponent, the corresponding amplitude appears to vanish for one or
more lattices and is non-zero otherwise. 
The idea of linking accidentally
(or otherwise) vanishing amplitudes to underlying
physical properties has been exploited fruitfully in the
past~\cite{bdn88}: the absence of vacancy corrections (with an exponent
$4/3$) has been demonstrated  for some Ising-like models, implying
completeness of the corresponding set of irrelevant operators;
investigation  also showed that this is particular to Ising systems, and
that for $q=2+ \epsilon$ Potts spins such
corrections are of order $\epsilon$.
In the present case, we
expect to probe the interplay between lattice symmetries
and the set of irrelevant operators for the respective spin
systems. To our knowledge, no attempt to draw such a connection has been  
made, despite the wealth of data available for two-dimensional
conformally invariant systems. 

\section{Ising ferromagnets}
\begin{figure}
\epsfxsize=16.5cm
\begin{center}
\epsffile{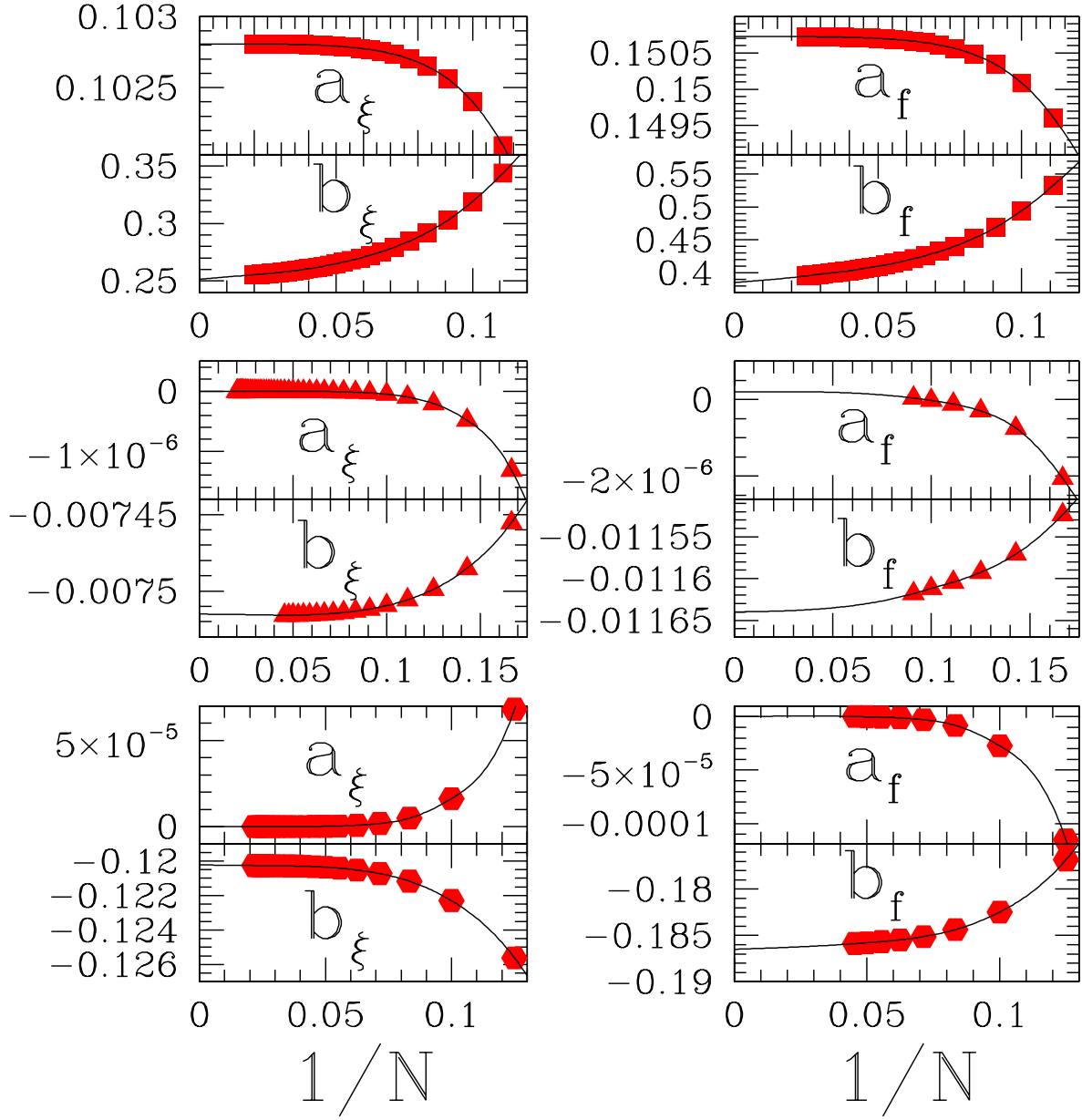}
\caption[]{
Two-point fits of coefficients in Equations (\protect\ref{eq:etai}) and 
(\protect\ref{eq:fei}). Splines are guides to the eye.
Lattices and
$N$-ranges displayed are (top to bottom) square: $a_{\xi}$, $b_{\xi}$:
$9-50$;  $a_f$, $b_f$: $9-40$; triangular: $a_{\xi}$: $6-27$; $b_{\xi}$:
$6-22$;  $a_f$, $b_f$: $6-11$; honeycomb: $a_{\xi}$, $b_{\xi}$:
$8-46$;  $a_f$, $b_f$: $8-22$ .
}
\label{fig:1}
\end{center}
\end{figure}

We start by considering the Ising model. Exact expressions
for the eigenvalues of the TM are available for all lattices 
concerned: square~\cite{dds,nig76,domb}, triangular~\cite{wannier}
and honeycomb~\cite{pf84,syozi}. From their inspection, one readily
sees that the finite-size estimates  $\xi_N$ and $f_N$ of
Equations (\ref{eq:1})
and (\ref{eq:2}) must have well-defined parity as functions of $N^{-1}$.
Therefore we assume, consistently with: (i) analytical evidence
derived for $\xi_N$ on square~\cite{dds} and honeycomb~\cite{pf84}
lattices (ii) numerical analysis of  $f_N$ on the square
lattice~\cite{bn}; (iii) the universal result connecting $f_N$ and conformal
anomaly~\cite{bcn}, and (iv) results for quantum Ising 
chains~\cite{malte,rein87}, that for all cases one has:
\begin{equation}
N/\pi\xi_N = 1/4 + a_{\xi}N^{-2}+ b_{\xi}N^{-4} + \ldots
\label{eq:etai}
\end{equation}
\begin{equation}
N^2(f_N- f_{\infty})= \pi/12 +
a_{f}N^{-2} +b_{f}N^{-4} + \ldots
\label{eq:fei}
\end{equation}
The  bulk free energies $f_{\infty}$ can be calculated in closed 
form~\cite{wu}, again for all three lattices. Truncating the series
above at $N^{-4}$, one gets finite-size approximants 
(also referred to as two-point fits) to the amplitudes 
$a_{\xi,f}$, $b_{\xi,f}$ from pairs of $\xi_N$ and $f_N$ for
consecutive  widths $N-1$ and $N$ ($N-2$ and $N$ for the honeycomb).
The approximants themselves 
still exhibit a weak $N$-dependence, on account of the
truncation just mentioned. In fact they usually converge rather smoothly
as $N$ increases, allowing reliable extrapolations to be produced; for
example, $a_{\xi}$ for the square lattice agrees with the exact
result~\cite{dds} to two parts in $10^5$.
We have found the sequences of two-point fits 
for $a_{\xi}$, $b_{\xi}$ to behave
better than those for $a_f$, $b_f$. For these latter on triangular
and honeycomb geometries, as well as $b_{\xi}$ on the triangular
lattice, numerical instabilities or sudden trend reversals
arose for large $N$; the worst
such cases were $a_f$, $b_f$ on the triangular lattice where behaviour
changed abruptly for $N > 11$.
We did not pursue the analysis of such deviations, since by then
we already had a fairly large sample of well-behaved data from which to
extrapolate (albeit with less accuracy than in other instances, where
monotonic trends seem to extend all the way as $N \to \infty$; for
instance, from a logarithmic plot one finds for the honeycomb
$a_{\xi}(N) \propto N^{-6.4}$ for $6 \leq N \leq 46$~). 
Graphical illustrations and extrapolated numerical estimates are shown
respectively in  Figure~\ref{fig:1} (where
use of $1/N$  on the horizontal axis is for ease of representation, not
implying assumption of a specific scaling form), and Table 1. 

\begin{table}
\caption{
Extrapolated ($N \to \infty$) amplitudes for Ising model 
(see Equations 
(\protect\ref{eq:etai}) and (\protect\ref{eq:fei})).
Uncertainties in last quoted digits shown in parentheses.
}
\begin{indented}
\item[]\begin{tabular}{@{}ccccc}
\br
\crule{5}\\
Lattice & $a_{\xi}$&$b_{\xi}$ &  $a_f$& $b_f$ \\
\mr
Square     &0.102810(2)$^{(\dagger)}$& 0.2515(5)& 0.150730(2)& 0.385(1)\\
Triangular & $< 10^{-8}$         
& $-0.007515(5)$&$ < 10^{-6}$ & $-0.01165(5)$\\
Honeycomb  & $< 10^{-10}$        & $-0.12022(5)$& $<10^{-6}$&$-0.1865(10)$\\
\br
\end{tabular}
\item $^{(\dagger)}$ Exact : $\pi^2/96 = 0.102808379 \ldots$
~\protect\cite{dds}
\end{indented}
\end{table}

The above results strongly suggest that the coefficients of the $N^{-2}$
corrections are exactly zero in both Equations (\ref{eq:etai}) and
(\ref{eq:fei}), for triangular and honeycomb geometries. 

\section{Three-state Potts ferromagnets}
In order to check whether the (apparently exact) vanishing of
the analytical corrections  is a model-independent, purely
lattice-related phenomenon, we proceeded to study the
next simplest spin system, the
three-state Potts ferromagnet. Exact
critical temperatures and exponents, and closed forms for
bulk free energies, are again
avaliable for all three lattices~\cite{wu}. Compared to the Ising case,
the main differences are: (i) no exact expressions are forthcoming for
the eigenvalues of the TM, implying that one must rely on
numerical diagonalisation, and also that the simple
argument for definite parity of eigenvalues as functions
of $N^{-1}$, plausible in the Ising case, need not apply here; and (ii)
(corroborating the point just made) a correction exponent $\omega_0=4/5$ 
is expected to arise~\cite{nienhuis}, overshadowing 
higher-order terms. 

We tackled (i) by generating the data displayed in Table 2,
for strip widths $3-14$ (square); $3-12$ (triangular) and $4-14$
(honeycomb; $N$ even). The  data in part (b) are displayed without the
geometric factor $\zeta$, in order to match the bulk free energies as
given in closed form in Ref.~\cite{wu}. Thus they must be multiplied
respectively by $2/\sqrt{3}$ (triangular) and $\sqrt{3}$ (honeycomb)
to fit \Eref{eq:2} with $c=4/5$.

\begin{table} 
\caption{
Finite-width data for 3-state Potts ferromagnet
}
\begin{indented}
\item[]\begin{tabular}{@{}llll}
\br
\ms
\centre{4}{ (a) $\eta_N = N/\pi\xi_N$}\\ 
$N$ & Square & Triangular& Honeycomb\\
\mr
 3 & 0.292\,265\,133\,729 & 0.274\,124\,359\,026 & \\
 4 & 0.282\,270\,742\,335 & 0.272\,011\,393\,185 & 0.271\,730\,743\,990\\
 5 & 0.277\,362\,861\,763 & 0.270\,966\,103\,989 & \\
 6 & 0.274\,717\,716\,167 & 0.270\,319\,649\,167 & 0.270\,584\,629\,588\\
 7 & 0.273\,119\,422\,538 & 0.269\,867\,269\,140 &  \\
 8 & 0.272\,058\,600\,969 & 0.269\,527\,830\,301 & 0.269\,822\,520\,229\\
 9 & 0.271\,304\,554\,006 & 0.269\,261\,539\,672 & \\
10 & 0.270\,741\,340\,874 & 0.269\,045\,974\,456 & 0.269\,315\,733\,294\\
11 & 0.270\,304\,823\,980 & 0.268\,867\,303\,915 & \\
12 & 0.269\,956\,670\,152 & 0.268\,716\,440\,914 & 0.268\,957\,780\,730\\
13 & 0.269\,672\,553\,348 &                      & \\
14 & 0.269\,436\,295\,211 &                      & 0.268\,691\,595\,289\\
$\infty^{(\dagger)}$ & 4/15 & 4/15 & 4/15 \\
\br
\ms
\centre{4}{ (b) $(1/N)\ln\Lambda_N^0$}\\ 
$N$ & Square$^{(\ast)}$ & Triangular& Honeycomb\\
\mr
 3 & 2.121\,091\,261\,980 & 2.002\,959\,900\,939 &  \\
 4 & 2.097\,704\,520\,030 & 1.985\,008\,921\,770 & 2.287\,656\,371\,786\\
 5 & 2.087\,460\,663\,806 & 1.976\,777\,367\,483 &  \\
 6 & 2.082\,063\,689\,903 & 1.972\,321\,291\,068 & 2.279\,250\,879\,301\\
 7 & 2.078\,863\,356\,335 & 1.969\,638\,661\,220 &  \\
 8 & 2.076\,806\,333\,203 & 1.967\,899\,052\,250 & 2.276\,305\,887\,939\\
 9 & 2.075\,404\,683\,690 & 1.966\,707\,034\,799 &  \\
10 & 2.074\,406\,246\,134 & 1.965\,854\,709\,111 & 2.274\,943\,228\,054\\
11 & 2.073\,669\,695\,186 & 1.965\,224\,253\,551 &  \\
12 & 2.073\,110\,712\,019 & 1.964\,744\,836\,946 & 2.274\,203\,278\,951\\
13 & 2.072\,676\,417\,341 &                      & \\
14 & 2.072\,332\,267\,371 &                      & 2.273\,757\,232\,635\\
$\infty^{(\ast \ast)}$ & 2.070\,187\,162\,576 & 1.962\,224\,155\,163 &
2.272\,522\,658\,739\\
\br
\end{tabular}
\item $^{(\dagger)}$ Conformal invariance~\protect\cite{cardy}
\item $^{(\ast)}$ Data for $N=3-11$ available in
Ref. \protect\cite{bn}
\item $^{(\ast \ast)}$ Evaluated in closed form~\protect\cite{wu}
\end{indented}
\end{table}

As regards (ii), we took recourse to predictions from conformal
invariance, namely: (a) analytical ($N^{-2}$) corrections are
always expected~\cite{cardynp}; (b) first-order corrections to the
ground-state free energy must be absent~\cite{car86,abb88}. Therefore we
assumed:
\begin{equation}
N/\pi\xi_N = 4/15 + a_{\xi}^{\omega_0}N^{-4/5}+
a_{\xi}^{2\omega_0}N^{-8/5} + a_{\xi}N^{-2} +a_{\xi}^{3\omega_0}N^{-12/5}
+ \ldots\ \  ;
\label{eq:etap}
\end{equation}
\begin{equation}
N^2(f_N -f_{\infty})=2\pi/15 +a_f^{2\omega_0}N^{-8/5} + 
a_{f}N^{-2} +a_f^{3\omega_0}N^{-12/5}+ \ldots\ \ .
\label{eq:fep}
\end{equation}
Because of the small value of $\omega_0$, second--, and even third--order
perturbation terms may give rise to corrections of comparable
magnitude to the analytical ones.
Obviously, one cannot simultaneously and reliably fit all coefficients
displayed in Equations~(\ref{eq:etap}) and (\ref{eq:fep}) from the data
of Table 2. We thus resorted to selective truncations. Recall that
our main goal is to check whether the presence or absence of analytical
corrections  follows the same lattice-dependent pattern found above.

For correlation-length data we started  by trying
(a1) two-point fits for $a_{\xi}^{\omega_0}$ and
$a_{\xi}^{2\omega_0}$, assuming $a_{\xi} \equiv 0$; 
(a2) same for $a_{\xi}^{\omega_0}$
and $a_{\xi}$, assuming $a_{\xi}^{2\omega_0} \equiv 0$; 
(b1) three-point fits 
for $a_{\xi}^{\omega_0}$, $a_{\xi}^{2\omega_0}$ and $a_{\xi}$;
(b2) three-point fits
for $a_{\xi}^{\omega_0}$, $a_{\xi}^{2\omega_0}$ and
$a_{\xi}^{3\omega_0}$,
assuming $a_{\xi}=0$. Least-squares
fits for varying ranges of $N$ were also performed to the
forms used in (a1)--(b2), always giving similar values of
$\chi^2 \sim 10^{-5} - 10^{-6}$ per degree of freedom. 
\begin{figure}
\epsfxsize=10.5cm
\begin{center}
\epsffile{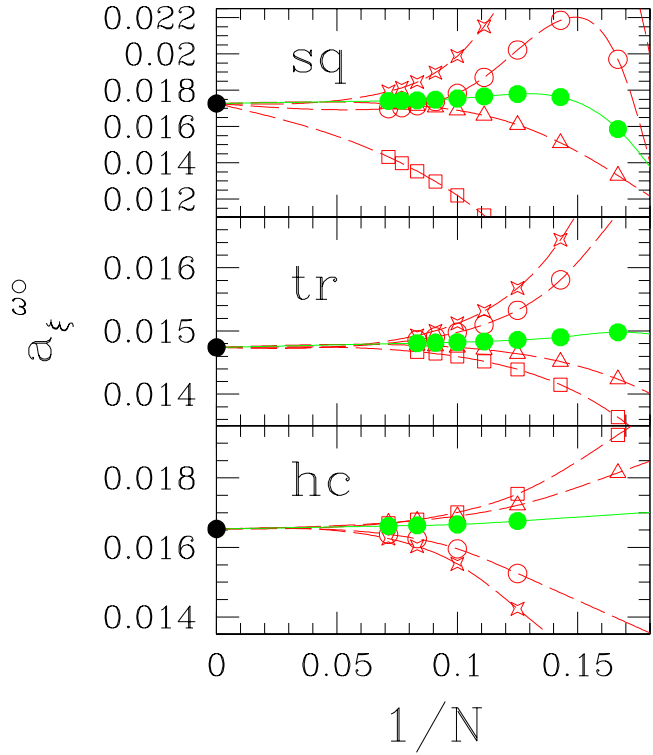}
\caption[]{
Finite-$N$ estimates of $a_{\xi}^{\omega_0}$  in Equation 
(\protect\ref{eq:etap}) from different truncation procedures
(see text). Squares: (a1); triangles: (a2); stars: (b1);
empty circles: (b2); full circles: Equation (\protect\ref{eq:etap2})
. 
Splines are guides to the eye, joining the vertical axis at
central estimates from Equation (\ref{eq:etap2}); see Table 3. 
Top to bottom: square, triangular, honeycomb.
}
\label{fig:2}
\end{center}
\end{figure}

For each lattice, all procedures gave reasonably consistent
trends for $a_{\xi}^{\omega_0}$,
no doubt because of the  small value of $\omega_0$, and the wide gap
separating it from all other
assumed secondary exponents; on the other hand, since these latter are 
so close to one another, we could generally extract no clear-cut
information from their respective fitted coefficients: in all cases,
any pair of terms (or single term)  among
$a_{\xi}^{2\omega_0}$, $a_{\xi}$, $a_{\xi}^{3\omega_0}$ could
do a reasonable job of standing in for effective corrections
when the other(s) was (were) assumed absent. Further, we noticed
that sequences of finite-$N$ estimates were much more stable than
any of the above if we used an {\it ad hoc} form 
inspired in the Ising case, namely  
\begin{equation}
N/\pi\xi_N = 4/15 + a_{\xi}^{\omega_0}N^{-4/5}+
a_{\xi}N^{-2} +b_{\xi}N^{-4}\ \ \ .
\label{eq:etap2}
\end{equation}
Indeed, in Figure~\ref{fig:2} one sees that, although for each lattice
all forms (a1)--(b2) yield estimates of $a_{\xi}^{\omega_0}$ whose
extrapolated values might conceivably coincide
(as illustrated by the splines),
 data from Equation~(\ref{eq:etap2})
exhibit the smallest residual $N$-dependence of all, leading to the least
uncertainties upon extrapolation. Such stability shows up also in the
estimates of $a_{\xi}$ of Equation~(\ref{eq:etap2}) and, to a more
limited extent, in those of $b_{\xi}$. This contrasts with the
corresponding sequences of $a_{\xi}^{2\omega_0}$ etc in procedures
(a1)--(b2), which generally  display a much broader variation. 
See Table 3 for the respective extrapolations  against $1/N$.

The small absolute values of $a_{\xi}$ for triangular and
honeycomb lattices remind one of the corresponding case for Ising 
spins. For (a2) on the triangular lattice, and (b1) on the
honeycomb,  error bars actually include the origin. 
However, turning now to the most regular series given by
Equation~(\ref{eq:etap2}), we recall that the respective estimates
are reached by crossing the horizontal axis, 
on extrapolation of rather
monotonic sequences; so it seems improbable that by  e. g. going to
larger widths a change of trend would occur, causing error bars
to be consistent with $a_{\xi}= 0$ in this case.
A different possibility, 
connected with the {\it ad hoc} character of Equation~(\ref{eq:etap2}),
is that there may be a {\it systematic} error implicit in assuming
this particular set of scaling corrections. One might ask
what combination, if any, of power-law corrections would be
consistent with the absence of analytical terms.
However, having already examined all reasonable forms suggested by theory
(namely procedures (a1)--(b2) and associated least-squares fits), and
having found none among them associated with clearly superior results,
as regards smooth convergence and/or quality of fit,
we decided not to proceed along these lines. 
\begin{table}
\caption{Extrapolated ($N \to \infty$) amplitudes for Potts correlation
lengths. See text for definitions of procedures (a1)--(b2), and
 Equation (\protect\ref{eq:etap2}).
Uncertainties in last quoted digits shown in parentheses.
}
\begin{indented}
\item[]\begin{tabular}{@{}llllll}
\br
\ms
Procedure& $a_{\xi}^{\omega_0} $& $a_{\xi}^{2\omega_0}$& $a_{\xi}$& 
$a_{\xi}^{3\omega_0}$& $b_{\xi} $ \\
\mr
\centre{6}{ Square }\\ \ms
(a1)     &  0.0189(7)  & 0.0352(6) & -- & -- & -- \\     
(a2)     &  0.0180(5) & -- & 0.119(6) & -- & -- \\     
(b1)     &  0.015(2) & 0.04(3) & 0.05(5) & -- & -- \\     
(b2)     &  0.016(1) & 0.038(15)& --& 0.13(5) & -- \\     
Eq. (\protect\ref{eq:etap2})     &  0.01727(4) &--& 0.133(1) &
-- &0.10(2)\\
\mr     
\centre{6}{ Triangular }\\ \ms
(a1)     &  0.0150(2)  & 0.000(1) & -- & -- & -- \\     
(a2)     &  0.0149(1) & -- & 0.001(1) & -- & -- \\     
(b1)     &  0.0141(6) & 0.012(7) & $-0.017$(10) & -- & -- \\     
(b2)     &  0.0144(3) & 0.006(3)& --& 0.012(6) & -- \\     
Eq. (\protect\ref{eq:etap2}) &  0.01474(1) & --& 0.0046(1)& 
-- & $-0.026(10)$ \\
\mr     
\centre{6}{ Honeycomb }\\ \ms
(a1)     &  0.0161(7)  & 0.0045(30) & -- & -- & -- \\     
(a2)     &  0.0163(4) & -- & 0.008(3) & -- & -- \\     
(b1)     &  0.0176(20) & $-0.016(25)$ & 0.028(31) & -- & -- \\     
(b2)     &  0.017(1) & $-0.005(9)$& --& 0.016(14) & -- \\     
Eq. (\protect\ref{eq:etap2}) &  0.01653(2) & -- & 0.0061(1)&
-- &$-0.31(2)$  \\
\br
\end{tabular}
\end{indented}
\end{table}

We now turn to free energies. Attempts to include a
first-order term, $a_f^{\omega_0}N^{-\omega_0}$ in
\Eref{eq:fep} consistently produced very small, and steadily
decreasing with increasing $N$, values of $a_f^{\omega_0}$ ($< 10^{-5}$,
on extrapolation) for all three lattices. Thus, our numerical data
are in entire accord with the prediction from conformal
invariance~\cite{car86,abb88} that $a_f^{\omega_0} \equiv 0$. 

Going to higher-order terms, we first recall earlier results, all for the
square lattice or, equivalently, quantum chains. 
A single-correction estimate gave $\omega=2$~\cite{bn}. Assuming
terms proportional to $N^{2\omega_0}$ to be negligible, contradictory
results were obtained upon comparing numerical estimates and
conformal-invariance predictions for gap amplitudes related to 
$N^{-2}$~\cite{vGRV}. This was later explained~\cite{rein87}
by showing that, though the amplitude of the  $N^{2\omega_0}$ terms
was indeed small, it could not be neglected. Specifically
as regards free energies, the following values 
were calculated, using the notation of our \Eref{eq:fep}:
$a_f^{2\omega_0} = 0.001721(3)$, $a_f = 0.0280(5)$~\cite{rein87}.

Again, we tried selective truncations of \Eref{eq:fep}, namely
(a1) two-point fits for $a_f^{2\omega_0}$ and
$a_f$, assuming $a_f^{3\omega_0} \equiv 0$; 
(a2) same for $a_f^{2\omega_0}$
and $a_f^{3\omega_0}$, assuming $a_f \equiv 0$; 
(b) three-point fits 
for $a_f^{2\omega_0}$, $a_f$ and $a_f^{3\omega_0}$.
Least-squares fits for varying ranges of $N$ were also performed to the
forms used in (a1)--(b). The results were generally
undistinguished, much as was the case for $\eta$ above. 
The only exception was (a2) for the
square lattice where, contrary to all other cases, we found a
steadily growing $a_f^{3\omega_0}$ with increasing $N$.
This clearly signals, consistently with earlier results~\cite{bn,rein87},
that  for the square lattice analytical terms are present with a large
coefficient.

In order to make contact with conformal invariance work, we investigated
the convergence of $a_f^{2\omega_0}$ and $a_f$ in procedure (a1), for the
square lattice. We found these estimates to vary significantly with $N$; 
in fact, Bulirsch-Stoer~\cite{bst,ms88} extrapolations pointed to
effective corrections $\sim N^{-x}$, $x \simeq 2.5$, with respective
final estimates $a_f^{2\omega_0} \simeq 0.02$, $a_f \simeq 0.25$. This
gives a value $\sim 12.5$ for the ratio $a_f/a_f^{2\omega_0}$, to be
compared to Reinicke's estimate, $16.3(3)$~\cite{rein87}. Given the
number and severity of approximations involved in our calculations,
this result may be regarded as broadly consistent with universality
of the ratio $a_f/a_f^{2\omega_0}$, as expected from conformal
invariance.

Since, for all procedures and lattices, the overall picture 
was very similar to that described in the preceding paragraph,
we decided to try the simpler scheme of Ref.~\cite{bn}, namely
assuming effective one-power
corrections $N^2(f_N -f_{\infty})
=2\pi/15 +a_{f}^{eff}N^{-\omega_0}$, and allowing $\omega_0$ to vary,
searching for good fits.
We determined the following optimum 
values for $\omega_0$: $1.9 - 2.0$, (square lattice, again in accord
with the early estimate  $\omega=2$~\cite{bn});
$1.6 - 1.7$ (triangular); $\sim 1.5$ (honeycomb). 
Similarly scattered  results arose recently from Monte-Carlo
calculations of magnetisation, susceptibility and specific heat for
three-state Potts spins on square $N \times N$ systems~\cite{kim},
where widely differing values of $\omega_0$
were obtained for the corrections of each quantity. As regards
the possible connection between lattice symmetry and presence,
or absence, of analytical corrections, the above one-power corrections
are broadly consistent with a similar picture to that
found for Ising systems. In such scenario, the triangular-
and honeycomb- effective exponents would reflect a non-zero coefficient
for $N^{-2\omega_0} =N^{-1.6}$, and a null one for the analytical term,
while for the square lattice the explanation would be as discussed
above: both terms are present, with $N^{-2}$ having an amplitude
one order of magnitude larger than $N^{-1.6}$.

\section{Conclusions} 
We have analysed  corrections to scaling in critical
Ising and three-state Potts ferromagnets on the three main 
two-dimensional lattices. By studying the simplest non-trivial systems 
available and  well-defined lattice symmetries, we have left aside
e.g. anisotropy-induced crossover phenomena and dealt
with power-law finite-size corrections alone, avoiding marginal operators
(which arise in four-state Potts systems, for instance) with their
associated logarithmic terms. We have given strong evidence that the
amplitudes of the $N^{-2}$ corrections of inverse correlation length and
free energy of Ising systems vanish both on triangular and honeycomb
geometries, but not on the square lattice. 

For the inverse correlation length of the Potts model, 
the main correction is well fitted by an $N^{-4/5}$ term, in accord
with theory and earlier numerical work~\cite{abb88,vGRV,rein87}. 
We have found the  values of the $N^{-2}$
coefficients for the triangular and honeycomb lattices to be certainly
much smaller than for a square geometry, though we have not been able to
ascertain that they vanish.

For Potts free energies, the $N^{-4/5}$ term is always absent, as
predicted by conformal invariance~\cite{car86,abb88}; we have estimated
the power arising in an effective single-term correction to be an
apparently lattice-dependent exponent in the range $1.5 - 2.0$.
Our results for the square lattice are consistent with earlier
work, in predicting $N^{-1.6}$ and $N^{-2}$ corrections whose
amplitude ratio is $\sim 12.5$, to be compared with the previously obtained
estimate $16.3(3)$ ~\cite{rein87}. As regards triangular
and honeycomb geometries, the respective exponents for the effective
single-term corrections are much closer to $2\omega_0 =8/5$ (the
second-order perturbation value predicted by conformal invariance) 
than to 2. This may, or may not, mean that the $N^{-2}$ term is absent in
these cases.

Several questions arise: 
(i) can one prove, e.g. on the basis of conformal
invariance properties,  that the amplitudes of the $N^{-2}$
corrections are exactly zero for triangular and honeycomb Ising systems?
If so, (ii) how do such amplitudes behave e.g.
in an anisotropic triangular
lattice as one crosses over towards a square symmetry, by reducing the
strength of the bonds along one of its main directions? And (iii) how
does the $2 + \epsilon$--state  Potts model behave in triangular or 
honeycomb geometries, as regards the $N^{-2}$ correction? 
(iv) Do the amplitudes of the $N^{-2}$ corrections really vanish  
for the three-state Potts model on the same lattices as it appears to be
the case for Ising systems? 
If so, can it be established by conformal invariance theory?
We plan to investigate numerical aspects of these and related matters   
in future work.

\ack
The author thanks: J A Castro for drawing his attention to the problem;
J Cardy for interesting conversations; Department of
Theoretical Physics at Oxford, where parts of this work were done, for
hospitality; the cooperation agreement between CNPq of Brazil
and the Royal Society for funding his visit;
CNPq, FINEP and CAPES for continuing financial support, and a referee
for drawing his attention to relevant references.


\end{document}